\documentclass[paper]{JHEP3}
\usepackage{cite}
\usepackage{amssymb,amsmath}
\usepackage{graphicx}
\def\bom#1{{\mbox{\boldmath $#1$}}}

\def\Q2{\left(Q^{2}\right)}
\def\e{\epsilon}

\def\CA{C_A}

\def\NF{N_F}

\def\MSbar{$\overline{{\rm MS}}$}

\def\l({\left(}
\def\r){\right)}

\def\e{\epsilon}

\def\sab{s_{12}}
\def\sac{s_{13}}
\def\sbc{s_{23}}

\def\sabc{s_{123}}

\newcommand{\be}{\begin{equation}}
\newcommand{\ee}{\end{equation}}
\newcommand{\bea}{\begin{eqnarray}}
\newcommand{\eea}{\end{eqnarray}}

\title{Two-loop QCD helicity amplitudes for $g\,g \to Z\,g$ and $g\,g \to Z\,\gamma  $}
\author{Thomas Gehrmann, Lorenzo Tancredi, Erich Weihs \\Institut f\"ur Theoretische Physik, 
Universit\"at Z\"urich, Wintherturerstrasse 190,\\CH-8057 Z\"urich, Switzerland}

\abstract{We compute the helicity amplitudes for the processes $gg\to Zg$ and 
$gg\to Z\gamma$ to two loops in massless QCD. The perturbative expansion 
of these processes starts only at the one-loop level, such that our results are a 
crucial ingredient to the NLO corrections to $Z\gamma$ and $Z+$jet production 
through gluon fusion.}
\preprint{{ZU-TH 04/13, LPN13-012}}
\begin{document}
% \begin{flushright}
% ZU-TH 13/13\\
% LPN13-12
% \end{flushright}

\maketitle
\allowdisplaybreaks

\section{Introduction}

The production of vector bosons at hadron colliders is, to a first 
approximation, induced by quark-antiquark annihilation. Including corrections from 
higher orders in the perturbative expansion in QCD, other processes will also contribute 
to vector boson final states. These contributions are suppressed by higher orders in 
the strong coupling constant $\alpha_s$, but could receive a numerical enhancement 
through the relevant parton-parton luminosity. In particular, 
in high-energy proton-proton collisions at the LHC, gluon-induced higher-order 
processes can become of comparable importance to quark-induced processes due 
to the large gluon luminosity at invariant masses relevant to vector boson production. 

Vector-boson production in gluon-gluon collisions is mediated through a quark loop, which 
vanishes for the exclusive $gg\to V$ vertex due to Furry's theorem. The gluon-gluon-induced 
subprocess becomes relevant for the production of vector boson pairs 
($WW$, $ZZ$, $\gamma \gamma$ and $Z\gamma$), or for the production of 
a neutral vector boson and a gluon. The leading-order scattering amplitudes for these 
processes all involve a closed quark loop. The resulting gluon-induced contributions 
from one-loop squared~\cite{costantini,baier,laursen} 
processes (that appear only at next-to-next-to-leading order in the
formal perturbative expansion of the full process) were evaluated a long time 
ago~\cite{glover,glover2,afs,bckk,diphox}, and typically found to yield a contribution 
that amounts to 10--20\% of the total cross section. Inclusion of 
these gluon-gluon subprocess contributions often results in an enhanced theoretical 
uncertainty on the prediction, since the one-loop squared process is effectively Born-level 
for this combination of partons. To stabilise these predictions, the computation of the next 
perturbative order in vector-boson pair production or vector-boson-plus-jet production in 
gluon fusion is required. Technically, such a calculation amounts to computing the 
corrections from single real radiation or single virtual exchange to the Born processes. With the 
Born process itself being a one-loop amplitude, one thus requires the two-loop corrections to 
the relevant partonic amplitudes. Up to now, these were obtained~\cite{twolgg} only for 
$gg\to \gamma\gamma$, where the NLO correction to the gluon-induced process 
was found to be sizeable and important in the stabilisation of the theoretical prediction for 
photon pair production~\cite{dixonschmidt,ggnnlo}.

In this paper, we derive
in massless QCD the two-loop corrections to the helicity amplitudes relevant to 
the production of a $Z$-boson in association with either a real photon or a hadronic jet 
in gluon-gluon collisions: $gg\to Z\gamma$ and $gg\to Zg$. For these processes, the 
one-loop amplitudes involving an extra gluon in the final state can be obtained using 
by-now standard methods for the computation of one-loop multi-leg 
processes~\cite{blackhat,rocket,gosam,cut tools,madloop,openloops}. With the results 
derived here, a complete NLO calculation of $Z\gamma$ and $Zj$ production in 
gluon fusion becomes thus feasible.

This paper is structured as follows: in Section~\ref{sec:kin}, we 
fix the notation and discuss the basic helicity structure of the process under
consideration. The general tensor structure of the amplitude is described in 
Section~\ref{sec:tensor} and expressed through helicity amplitudes in Section~\ref{sec:hel}.
The calculation of the two-loop amplitudes, their renormalisation and 
infrared properties and their simplification are  
described in Section~\ref{sec:calc}. 
The two-loop helicity amplitudes are 
obtained in a closed analytic form.
We performed several non-trivial checks on the results,
which are described in Section~\ref{sec:checks}. We conclude with an 
outlook in Section~\ref{sec:conc}.
We enclose appendices with the analytical form for the one-loop and two-loop
helicity amplitudes in the decay kinematics $V \to ggg$ and $V \to gg\gamma$.
The helicity amplitudes continued to the regions relevant for 
vector-boson-plus-jet and vector-boson-plus-photon production at LHC are given 
in Mathematica format together with the arXiv submission of this paper.

%\section{Kinematics and basic helicity structure}
%\label{sec:kin}

\section{Kinematics and notations}
\label{sec:kin}
The production of a massive vector boson $ V = (Z^0,\gamma^*)$ and a gluon (photon) through gluon-gluon fusion is related by crossing
to the decay of a massive vector boson to three gluons (two gluons and a photon) and has the same kinematics
as vector-boson-plus-jet production $q \bar{q} \to V g$, $q g \to V q$ and vector-boson-plus-photon production $q \bar{q} \to V \gamma$.
Technically the calculation of the two-loop QCD corrections to the $gg\to Vg$ and $gg \to V \gamma $ amplitudes is thus 
similar to previous calculations for $3j$-production, vector-boson-plus-photon production and
 $H\to 3$~partons,  
which have been derived to two-loop accuracy in QCD~\cite{3jtensor,Vgamma,Hggg}.

In the following we will focus on the decay kinematics, while the crossings relevant for $V$-plus-jet and $V$-plus-photon
production at hadron colliders will be discussed in  section~\ref{sec:crossings}.

The relevant partonic subprocesses are:
\begin{eqnarray}
&& l^-(p_5) + l^+(p_6) \to V(q) \to g(p_1) + g(p_2) + g(p_3) \;, \nonumber \\
&& l^-(p_5) + l^+(p_6) \to V(q) \to g(p_1) + g(p_2) + \gamma(p_3)\; , \label{1A}\;
\end{eqnarray}
where we included the production of the vector boson $V$ through lepton-antilepton annihilation.

In the framework of massless QCD interchanging the virtual photon 
with a $Z$ boson amounts only to a proper re-weighting of the final result.
Moreover, note that we always assume massless fermions in the initial or final state.

The momentum of the vector boson is given by
\begin{equation}
q^{\mu} = p_1^{\mu} +p_2^{\mu} +p_3^{\mu}\;.
\end{equation}
It is convenient to define the usual invariants
\begin{equation}
\sab = (p_1+p_2)^2\;, \qquad \sac = (p_1+p_3)^2\;, \qquad 
\sbc = (p_2+p_3)^2\;,
\end{equation}
which fulfil
\begin{equation}
q^2  =(p_1+p_2+p_3)^2 = \sab + \sac + \sbc \equiv \sabc \; ,
\end{equation}
as well as the dimensionless invariants
\begin{equation}
x = \sab/\sabc\;, \qquad y = \sac/\sabc\;, \qquad z = \sbc/\sabc\;,
\end{equation}
which satisfy $x+y+z=1$.

In the decay kinematics $V \to ggg /gg\gamma $, as in the $3j$ case, $q^2$ is time-like (hence 
positive) and all the $s_{ij}$ are also positive, which implies that 
$x,y,z$ all lie in the interval $[0;1]$, with the above constraint $x+y+z=1$.

The helicity amplitudes can be expressed as a product of a partonic current $S_\mu$ and a leptonic current $L_\mu$:
\begin{align}
 A(p_5,p_6;g_1,g_2,b_3) &= L^\mu(p_5;p_6) S_\mu(g_1;g_2;b_3)\;
\end{align}
where $g_i = g(p_i)$, and $b_3 = b(p_3)$ labels a generic massless gauge boson. 
In our case $b = g,\gamma$ in $V \to ggg$ and $ V \to gg \gamma$ respectively.\newline

The purely vectorial tree-level leptonic current reads:
\begin{equation}\label{eq_leptoniccurrent}
L^{\mu}(p_5,p_6) = \bar{v}(p_6) \, \gamma^{\mu} \, u(p_5),
\end{equation}
where in the case of an incoming lepton-antilepton pair $L_\mu(p_5^-,p_6^+)$ 
corresponds to a left-handed current, and $L_\mu(p_5^+,p_6^-)$ to a 
right-handed current: 
\be
L_L^\mu(p_5^-,p_6^+) = \bar{v}_+(p_6) \, \gamma^{\mu} \, u_-(p_5), \qquad
L_R^\mu(p_5^+,p_6^-) = \bar{v}_-(p_6) \, \gamma^{\mu} \, u_+(p_5).
\ee
Only the partonic currents receive contributions from QCD radiative corrections,
and they can be perturbatively decomposed as:
\bea
 S_\mu(g_1;g_2;g_3) 
 =&  \sqrt{4 \pi \alpha_s}\,d^{a_1 a_2 a_3} \, \Big[ 
 \left(\frac{\alpha_s}{2\pi}\right)    S^{(1)}_\mu(g_1;g_2;g_3)\nonumber 
+ \left(\frac{\alpha_s}{2\pi}\right)^2 S^{(2)}_\mu(g_1;g_2;g_3) 
+ {\cal O}(\alpha_s^3) \, \Big]\;,\\
 S_\mu(g_1;g_2;\gamma_3) 
 =&  \sqrt{4 \pi \alpha} \,\delta^{a_1 a_2} \, \Big[ 
 \left(\frac{\alpha_s}{2\pi}\right)    S^{(1)}_\mu(g_1;g_2;\gamma_3)\nonumber 
+ \left(\frac{\alpha_s}{2\pi}\right)^2 S^{(2)}_\mu(g_1;g_2;\gamma_3) 
+ {\cal O}(\alpha_s^3) \, \Big]\;,
\eea
where we factored out the overall colour factors $\delta^{a_1a_2}$, $d^{a_1 a_2 a_3}$.

The general form of the gauge boson coupling to fermions is:
\begin{equation}
\mathcal{V}_\mu^{V,f_1f_2} = - i \,e\, \Gamma_{\mu}^{V,f_1f_2}  \qquad \mbox{with} \qquad e = \sqrt{4 \pi \alpha},
\end{equation}
whose explicit form depends on the gauge boson, on the type of 
fermions, and on their helicities:
\begin{equation}
\Gamma_{\mu}^{V,f_1f_2} = L_{f_1f_2}^V \; \gamma_\mu \left(\frac{1-\gamma_5}{2}\right) 
                     + R_{f_1f_2}^V \; \gamma_\mu \left(\frac{1+\gamma_5}{2}\right). 
\end{equation}
The left- and right-handed couplings are identical for a pure vector interaction, and 
are in general different if vector and axial-vector interactions contribute.
Their values for a photon are
\begin{equation}
L_{f_1f_2}^{\gamma^*} = R_{f_1f_2}^{\gamma^*} = -e_{f_1}\, \delta_{f_1f_2},
\end{equation}
while for a $Z$ boson
\begin{equation}
L_{f_1f_2}^Z = \frac{I_3^{f_1} - \sin^2{\theta_w} e_{f_1}}{\sin{\theta_w} \cos{\theta_w}}\, \delta_{f_1f_2}\;,\qquad \qquad 
R_{f_1f_2}^Z = -\frac{\sin{\theta_w} e_{f_1}}{ \cos{\theta_w}}\, \delta_{f_1f_2}\;.
\end{equation}

The vector boson propagator can be written as:
\begin{align}
P_{\mu \nu}^V(q,\xi) &= \frac{ i\, \Delta_{\mu \nu}^V(q,\xi) }{D_V(q)},
\end{align}
where  $\Delta_{\mu \nu}^{V}(q,\xi)$ and $D_V(q)$ are, respectively, the numerator 
and the denominator in the $R_\xi$ gauge:
\begin{align}
&\Delta_{\mu\nu}^{V}(q,\xi) = \left(-g_{\mu\nu} + (1-\xi) \frac{q_\mu q_\nu}{q^2-\xi M_V^2} \right),\label{numProp}
\end{align}
\begin{eqnarray}
 D_{Z}(q) &=& \big(\, q^2 - M_Z^2 + i \Gamma_Z M_Z \, \big), \label{denProp} \\
 D_{\gamma^*}(q)  &=& q^2.  
\end{eqnarray}

In the narrow-width approximation we can simplify expression~\eqref{denProp} to 
\begin{equation}
D_{Z}(q) \approx i \Gamma_Z M_Z  \qquad \mbox{and} \qquad q^2 = M_Z^2,
\end{equation}
where $M_Z$ is the mass of the $Z$ boson, while
$\Gamma_Z$ is its decay width. 

Since we do not consider any electroweak corrections, the vector boson $V$
is always coupled to a fermion line which allows us to neglect the $R_\xi$ dependence 
(or equivalently to put $\xi=1$). A further consequence is that the 
 total amplitude is  proportional to the charge weighted sum over the quark flavours, such that 
all electroweak couplings can be collected into a multiplicative factor
$Q_V^{b}$. 
With this notation we obtain for an incoming right-handed lepton-antilepton pair, for the different 
choices of $V = (\gamma^*,Z)$, and helicity configurations $(h_1,\,h_2,\,h_3)$:
\begin{align}
\mathcal{M}_V(p_5^+,p_6^-;g_1^{h_1},g_2^{h_2},b_3^{h_3}) &=
-i\,(4 \pi \alpha)\, \frac{\,R_{f_5f_6}^V\,Q_V^b}{D_V(p_{5}+p_{6})}\, A_R^{(h_1\,h_2\,h_3)}(p_5,p_6;g_1,g_2,b_3),
\end{align}
In case of $V=\gamma^*$ we find
\bea
&Q_{\gamma^*}^g = \sum_{q} e_{q},\\
&Q_{\gamma^*}^\gamma =\sum_{q} e^2_{q},
\eea
where the sum runs over the quark flavours in the loop.

In the case of $V=Z$ we have a contribution from the vector component of the $Z$ boson, which is given by
\bea 
&Q_{Z}^g  = \frac{1}{2}\sum_{q} (L_{qq}^Z + R_{qq}^Z),\\
&Q_{Z}^\gamma  = \frac{1}{2}\sum_{q} (L_{qq}^Z + R_{qq}^Z)e_{q},
\eea 
but also a contribution involving its axial coupling. This contribution vanishes 
for $Z\to gg \gamma$ due to charge conjugation invariance,
 already before summing over the quark flavours in the loop.
On the other hand, in the case of $Z \to ggg$ it vanishes only after
summing over the quark flavours.

\section{The general tensor structure}
\label{sec:tensor}
In order to extract the helicity amplitudes from a generic QCD process different approaches can be attempted.
One possibility is to decompose the amplitude into linearly independent tensor structures,
whose number and form are entirely determined by symmetry considerations and which 
are completely independent
on the loop order we are interested in.
The entire loop-dependence is then contained in the scalar coefficients which multiply the relevant tensor structures.
In order to single out these coefficients we apply projectors defined in $d$-continuous dimensions directly on
the Feynman-diagrammatic expression for the amplitude~\cite{3jtensor,Vgamma,Hggg,SusyQED}.

Using Lorentz invariance one can show that there are 138 independent Lorentz structures which can contribute to the partonic current~\cite{SusyQED}:
\bea\label{tensordec}
S^{\mu \nu\rho\sigma} &=& 
        a_1 g^{\mu\nu} g^{\rho\sigma}
       +a_2 g^{\mu\rho} g^{\nu\sigma}  
       +a_3 g^{\mu\sigma} g^{\nu\rho} \nonumber \\
&&+\sum\limits_{j_1,j_2=1}^{3} 
\Bigl(
   b^1_{j_1 j_2}\,g^{\mu\nu}\,p_{j_1}^{\rho}\,p_{j_2}^{\sigma}
  +b^2_{j_1 j_2}\,g^{\mu\rho}\,p_{j_1}^{\nu}\,p_{j_2}^{\sigma}
  +b^3_{j_1 j_2}\,g^{\mu\sigma}\,p_{j_1}^{\nu}\,p_{j_2}^{\rho}\nonumber \\ 
&&\phantom{\sum\limits_{j_1,j_2=1}^{3}}
+b^4_{j_1 j_2}\,g^{\nu\rho}\,p_{j_1}^{\mu}\,p_{j_2}^{\sigma}
  +b^5_{j_1 j_2}\,g^{\nu\sigma}\,p_{j_1}^{\mu}\,p_{j_2}^{\rho} 
  +b^6_{j_1 j_2}\,g^{\rho\sigma}\,p_{j_1}^{\mu}\,p_{j_2}^{\nu}
\Bigr) 
\nonumber \\  
 &&+\sum\limits_{j_1,j_2,j_3,j_4=1}^{3} c_{j_1 j_2 j_3 j_4}
p_{j_1}^{\mu}\,p_{j_2}^{\nu} p_{j_3}^{\rho}\,p_{j_4}^{\sigma}.
\eea
Not all these tensors will be relevant for our computations.
Defining the physical amplitude contracted with the external polatization vectors of the three massless on-shell bosons:
\begin{align} 
 S_{\mu}(g_1;g_2;b_3) &= S_{\mu \nu \rho \sigma}(p_1;p_2;p_3)\,\epsilon_1^\nu(g)\, \epsilon_2^\rho(g)\,  \epsilon_3^\sigma(b),
% S_{\mu}(g_1;g_2;g_3) &= S_{\mu \nu \rho \sigma}(p_1;p_2;p_3)\,\epsilon_1^\nu\, \epsilon_2^\rho\,  \epsilon_3^\sigma,\\
% S_{\mu}(g_1;g_2;\gamma_3) &= S_{\mu \nu \rho \sigma}(p_1;p_2;p_3)\,\epsilon_1^\nu\, \epsilon_2^\rho\,  \epsilon_3^\sigma,
\end{align}
we see that many of the structures do not contribute because of the transversality condition:
$$\epsilon_i \cdot p_i = 0, \qquad \mbox{with} \quad i=1,2,3\,. $$
This reduces the number of independent tensors to 57.
One way of proceeding is then to apply Ward identities for the massless bosons
\begin{align} 
S^{\mu \nu \rho \sigma}\, p_1^\nu\, \epsilon_2^\rho\, \epsilon_3^\sigma = 
S^{\mu \nu \rho \sigma}\, \epsilon_1^\nu\, p_2^\rho\, \epsilon_3^\sigma = 
S^{\mu \nu \rho \sigma}\, \epsilon_1^\nu\, \epsilon_2^\rho\, p_3^\sigma = 0 \label{WIs}.
\end{align}
which lowers the number of relevant structures down to 18.
Applying finally current conservation for the massive boson
\begin{align} 
&S^{\mu \nu \rho \sigma}\, \epsilon_1^\nu\, \epsilon_2^\rho\, \epsilon_3^\sigma\, p_4^\mu = 0 \label{Curr}.
\end{align}
further reduces the number of independent tensor coefficients to 14. 

%As shown in section~\ref{sec:checks}, 
By requiring the amplitude to be invariant under the exchange 
of the three (two) gluons one can find further relations 
among these 14 coefficients with interchanged arguments. 
This allows to perform different checks on the final result (see section~\ref{sec:checks}).

Once the tensor structure is known, one can compute $d$-dimensional projection operators 
that applied on $S_{\mu \nu \rho \sigma}$ extract each of the 14 coefficients. 
The tensors and the projectors contain a large number of individual terms. 
Therefore applying them to an amplitude in a Feynman-diagrammatic approach 
will in general result in a large number of contractions with a huge
proliferation of terms.

Moreover, it must be noted that the basis of tensors is not unique, 
namely that any set of 14 tensors, obtained as independent linear combinations of those found above, can be chosen.
Choosing suitable linear combinations of the above tensors can simplify their structure 
substantially.

For all these reasons we decided to follow a simplified approch, 
which nevertheless allows us to retain the full information on the process.
It is well known that when performing a computation with a large number of external bosons
a specific gauge choice can highly simplify the intermediate steps of the calculation,
while gauge invariance ensures that the final result for the amplitude must be independent on the choice made.
Following this idea, instead of imposing gauge invariance on the tensor structures, 
we chose to fix the gauge of the external particles in order to symplify the tensor structures as much as possible.

Naively one would expect the loss of gauge invariance on the tensors, 
together with the loss of part of the symmetry due to the gauge choice performed,
to be a drawback of this approach. 
However, one can show that once these 14 coefficients are known, the full gauge-invariant tensor
can be reconstructed. 
In particular one can find linear relations among the 14 
coefficients obtained imposing the gauge fixing and the 14 coefficients of the gauge invariant tensor, 
as outlined in the following section.

\subsection{The gauge-fixed tensor structure}
Following the above reasoning, 
 we replace the condition~\eqref{WIs} with a gauge choice on the external on-shell bosons:
\begin{equation}
\epsilon_1\cdot p_2 = \epsilon_2 \cdot p_3 = \epsilon_3 \cdot p_1 = 0. \label{Gauge}
\end{equation}
This choice is arbitrary and could be substituted by any other set of gauge conditions.
The advantage of this particular choice is to produce extremely compact tensor structures.

Fixing the gauge of the external bosons reduces the number of independent tensors to 18.
Also in this case we impose current conservation~\eqref{Curr} on the $Z^0$ 
and end up again with 14 tensor structures. 
As expected, the number of independent tensor structures obtained in this way is the same
as for the gauge-independent tensor. 

We decompose the parton current as
\be
S^{\mu}(g_1,g_2,b_3) = \sum_{i=1}^{14} A_{i}^{(b)} \, T_{i}^\mu\,, \label{eq:smubase}
\ee
where the coefficients are functions of the mandelstam variables $A_i^{(b)} = A_i^{(b)}(\sab,\sac,\sbc)$
and their explicit values differ in general if $b$ is a gluon or a photon.

Finally, the gauge-fixed tensors read:
\begin{align}
T_{1}^\mu &= \epsilon_1 \cdot p_3\, \epsilon_3 \cdot p_2 \, \epsilon_2^\mu
           - \epsilon_2 \cdot p_1\, \epsilon_3 \cdot p_2 \, \epsilon_1^\mu, \qquad
T_{2}^\mu = \epsilon_1 \cdot p_3\, \epsilon_2 \cdot p_1 \, \epsilon_3^\mu
           - \epsilon_2 \cdot p_1\, \epsilon_3 \cdot p_2 \, \epsilon_1^\mu,
\end{align}

\begin{align}
T_{3}^\mu &= \epsilon_1 \cdot \epsilon_2 \, \left[ \epsilon_3\cdot p_2  \, p_1^\mu 
                                                 - {( s_{12}+s_{13} ) \over 2} \,{\epsilon_3^\mu } \right]\,,\\
T_{4}^\mu &= \epsilon_1 \cdot \epsilon_2 \, \left[ \epsilon_3\cdot p_2  \, p_2^\mu 
                                                 - {( s_{12}+s_{23} ) \over 2} \,{\epsilon_3^\mu } \right]\,,\\
T_{5}^\mu &= \epsilon_1 \cdot \epsilon_2 \, \left[ \epsilon_3\cdot p_2  \, p_3^\mu 
                                                 - {( s_{13}+s_{23} ) \over 2} \,{\epsilon_3^\mu } \right]\,,
\end{align}

\begin{align}
T_{6}^\mu &= \epsilon_1 \cdot \epsilon_3 \, \left[ \epsilon_2\cdot p_1  \, p_1^\mu 
                                                 - {( s_{12}+s_{13} ) \over 2} \,{\epsilon_2^\mu } \right]\,,\\
T_{7}^\mu &= \epsilon_1 \cdot \epsilon_3 \, \left[ \epsilon_2\cdot p_1  \, p_2^\mu 
                                                 - {( s_{12}+s_{23} ) \over 2} \,{\epsilon_2^\mu } \right]\,,\\
T_{8}^\mu &= \epsilon_1 \cdot \epsilon_3 \, \left[ \epsilon_2\cdot p_1  \, p_3^\mu 
                                                 - {( s_{13}+s_{23} ) \over 2} \,{\epsilon_2^\mu } \right]\,,
\end{align}

\begin{align}
T_{9}^\mu &= \epsilon_2 \cdot \epsilon_3 \, \left[ \epsilon_1\cdot p_3  \, p_1^\mu 
                                                 - {( s_{12}+s_{13} ) \over 2} \,{\epsilon_1^\mu } \right]\,,\\
T_{10}^\mu &= \epsilon_2 \cdot \epsilon_3 \, \left[ \epsilon_1\cdot p_3  \, p_2^\mu 
                                                 - {( s_{12}+s_{23} ) \over 2} \,{\epsilon_1^\mu } \right]\,,\\
T_{11}^\mu &= \epsilon_2 \cdot \epsilon_3 \, \left[ \epsilon_1\cdot p_3  \, p_3^\mu 
                                                 - {( s_{13}+s_{23} ) \over 2} \,{\epsilon_1^\mu } \right]\,,
\end{align}

\begin{align}
T_{12}^\mu &= \epsilon_2 \cdot p_1\, \epsilon_3 \cdot p_2 \, \left[ \epsilon_1\cdot p_3  \, p_1^\mu 
                                                 - {( s_{12}+s_{13} ) \over 2} \,{\epsilon_1^\mu } \right]\,,\\
T_{13}^\mu &= \epsilon_2 \cdot p_1\, \epsilon_3 \cdot p_2 \, \left[ \epsilon_1\cdot p_3  \, p_2^\mu 
                                                 - {( s_{12}+s_{23} ) \over 2} \,{\epsilon_1^\mu } \right]\,,\\
T_{14}^\mu &= \epsilon_2 \cdot p_1\, \epsilon_3 \cdot p_2 \, \left[ \epsilon_1\cdot p_3  \, p_3^\mu 
                                                 - {( s_{13}+s_{23} ) \over 2} \,{\epsilon_1^\mu } \right]\,.
\end{align}

The relations among the $A_i^{(b)}$ and the coefficients of the gauge invariant tensor can be found by 
performing on the latter the gauge fixing~\eqref{Gauge}.
This procedure obviously does not affect the scalar coefficients which multiply the tensor structures. 
One ends up then with 14 new tensor structures which can be related through linear combinations
to those obtained fixing the gauge from the beginning. In this way the gauge invariant tensor can be fully reconstructed.
We have verified this procedure by comparing our one-loop result with the literature \cite{glover} 
where the results are given for an on-shell $Z$ boson, and a different gauge choice is used (see section \ref{sec:checks}).

Once the tensor structure is known, one can obtain the coefficients $A_{i}^{(b)}$ by applying a set of projectors $P_\mu(A_{i}^{(b)})$ 
on the Feynman-diagrammatic expression of the amplitude defined such that $$\sum_{spin} P^\mu(A_{i}^{(b)}) \, S_\mu(p_1,p_2,p_3) = A_{i}^{(b)}.$$

Note that the projection has to be performed in $d$ dimensions,
and that special care has to be taken in performing the polarization sums when applying the projectors on the single diagrams. 
In particular one has to consistently use a physical polarization sum which respects the gauge choice~\eqref{Gauge}:

\begin{align}
 \sum_{spin} \epsilon_{1}^{*\mu}(p_1)\, \epsilon_1^\nu(p_1) &= - g^{\mu \nu} + \frac{p_1^\mu \, p_2^\nu + p_1^\nu \, p_2^\mu }{ p_1 \cdot p_2},\\
 \sum_{spin} \epsilon_{2}^{*\mu}(p_2)\, \epsilon_2^\nu(p_2) &= - g^{\mu \nu} + \frac{p_2^\mu \, p_3^\nu + p_2^\nu \, p_3^\mu }{ p_2 \cdot p_3},\\
 \sum_{spin} \epsilon_{3}^{*\mu}(p_3)\, \epsilon_3^\nu(p_3) &= - g^{\mu \nu} + \frac{p_3^\mu \, p_1^\nu + p_3^\nu \, p_1^\mu }{ p_3 \cdot p_1}.
\end{align}

The projectors themselves can be decomposed in the tensor basis and take the form:
\be
P^\mu(A_{j}^{(b)}) = \sum_{j=1}^{14} X_{i}(A_{j}^{(b)})\,T_{i}^{*\mu}
\ee
where the $X_{i}(A_j^{(b)})$ are functions of $d$ and the kinematical invariants $s_{ij}$. 
% Each of the unrenormalised coefficients $A_{i}$ has a perturbative expansion of the form
% \begin{eqnarray}
% A_{i}^{{\rm un}} &=& \sqrt{4 \pi \alpha_s}\,d^{abc}\left[ 
%   \left(\frac{\alpha_s}{2\pi}\right) A_{i}^{(1),{\rm un}}  
% + \left(\frac{\alpha_s}{2\pi}\right)^2 A_{i}^{(2),{\rm un}} 
% + {\cal O}(\alpha_s^3) \right] \;,\nonumber 
% \end{eqnarray}
% where the dependence on $(u,v)$ is now implicit.

\section{Helicity amplitudes}
\label{sec:hel}
By fixing the helicities of the external massless bosons the partonic current can be cast in the usual spinor helicity notation~\cite{dixon}.
There are two independent helicity configurations in the $ggg V$-case, and three independent helicity configurations in the $gg\gamma V$-case, 
from which all the others can be obtained.
In the following we discuss separately the two cases.

\subsection{$ggg V$: The amplitude in spinor helicity notation}
We start off considering the $ggg V$-case. 
We choose as two independent helicity configurations $(g_1^+,g_2^-,g_3^-)$ and $(g_1^+,g_2^+,g_3^+)$.
In order to include the spin-correlations with the leptonic decay products 
we contract the partonic current with the leptonic current $L_\mu$ for fixed helicities of the
initial state leptons. This also helps to further simplify the result.

Consider the production of the vector boson $V$ through lepton-antilepton  annihilation:

$$ l^-(p_5) + l^+(p_6) \longrightarrow V(q).$$
The leptonic currents (\ref{eq_leptoniccurrent}) are
\begin{equation}
L_{R}^\mu(p_5^+,p_6^-) =  [6 \;| \gamma^\mu |\; 5 \rangle , \qquad
L_{L}^\mu(p_5^-,p_6^+) =  [5 \;| \gamma^\mu |\; 6 \rangle = [L_{R}^\mu(p_5^+,p_6^-)]^*.
\end{equation}

Performing the contraction and making use of Schouten identities and momentum conservation we end up with:
\begin{align}
A_R^{(+--)}&(p_5,p_6;g_1,g_2,g_3) = L_R^\mu(p_5^+,p_6^-) \, S_\mu(g_1^+,g_2^-,g_3^-) = 
 \frac{1}{\sqrt{2}} \, \frac{\langle 2\,3\rangle }{\langle 1\, 2\rangle \langle 1\, 3\rangle  [2\,3]} \nonumber \\ \times  \Bigg\{ 
& \langle 2\,5\rangle \langle 3\,5\rangle  [6\,5] \, \alpha_1(x,y,z) 
+ \langle 2\,3\rangle \langle 2\,5\rangle  [2\,6] \, \alpha_2(x,y,z)
+ \langle 2\,3\rangle \langle 3\,5\rangle  [3\,6] \, \alpha_3(x,y,z) \, \Bigg\}\;, \label{PMMg}
\end{align}

\begin{align}
 A_R^{(+++)}&(p_5,p_6;g_1,g_2,g_3) = L_R^\mu(p_5^+,p_6^-) \, S_\mu(g_1^+,g_2^+,g_3^+) =
 \frac{1}{\sqrt{2}} \nonumber \\\times  \Bigg\{ 
&\frac{[1\,3] \langle 1\,5\rangle  [1\,6]}{\langle 1\, 2\rangle \langle 2\,3\rangle} \beta_1(x,y,z)
+\frac{[2\,3] \langle 2\,5\rangle  [2\,6]}{\langle 1\, 2\rangle \langle 1\,3\rangle} \beta_2(x,y,z)
+\frac{[2\,3] \langle 2\,5\rangle  [1\,6]}{\langle 1\, 2\rangle \langle 2\,3\rangle} \beta_3(x,y,z)    \Bigg\}\;, \label{PPPg}
\end{align}
where the coefficients $\alpha_i$ and $\beta_i$ are linear combinations of the 14 tensor coefficients $A_i$.
As an explicit example we write down the relations for the $\alpha_j$:
\begin{align}
\alpha_1(x,y,z) &= -(s_{12}+s_{13}) \left[ A_2 + A_9 + {s_{12} \over 2 } A_{12} \right] 
                 -(s_{12}+s_{23}) \left[ A_1 + A_{10} + {s_{12} \over 2 } A_{13} \right] \nonumber \\
                &\quad -(s_{13}+s_{23}) \left[ A_{11} + {s_{12} \over 2 } A_{14} \right],  \\
\alpha_2(x,y,z) &= - s_{12}\left[ A_2 + A_9 + {s_{12} \over 2 } A_{12} \right]
                 - (s_{12}+s_{13}+s_{23})\left[ A_1 + A_{10} + {s_{12} \over 2 } A_{13} \right] \nonumber \\
                &\quad -(s_{13}+s_{23}) \left[ A_{11} + {s_{12} \over 2 } A_{14} \right],  \\
\alpha_3(x,y,z) &=  s_{13} \left[ A_{2} + A_9 - A_{11} +  {s_{12} \over 2 } A_{12}  - {s_{12} \over 2 } A_{14} \right].
\end{align}
The corresponding relations for the $\beta_j$ are slightly longer and we do not
 reproduce them here for brevity. 
There are in total 16 different helicity configurations. From the above expressions for 
$A_R^{(+--)}(p_5,p_6;g_1,g_2,g_3)$ and $A_R^{(+++)}(p_5,p_6;g_1,g_2,g_3)$, all the other helicity amplitudes can be obtained
by parity conjugation and permutations of the external legs. We find:
\begin{align}
 L_R^\mu(p_5^+,p_6^-) \, S_\mu(g_1^-,g_2^+,g_3^-) &= A_R^{(-+-)}(p_5,p_6;g_1,g_2,g_3) = A_R^{(+--)}(p_5,p_6;g_2,g_1,g_3)\,,   \nonumber\\
 L_R^\mu(p_5^+,p_6^-) \, S_\mu(g_1^-,g_2^-,g_3^+) &= A_R^{(--+)}(p_5,p_6;g_1,g_2,g_3) = A_R^{(+--)}(p_5,p_6;g_3,g_2,g_1)  \,, \nonumber\\
%%%%%
 L_R^\mu(p_5^+,p_6^-) \, S_\mu(g_1^+,g_2^+,g_3^-) &= A_R^{(++-)}(p_5,p_6;g_1,g_2,g_3) = [A_R^{(+--)}(p_6,p_5;g_3,g_2,g_1)]^*\,,  \nonumber\\
 L_R^\mu(p_5^+,p_6^-) \, S_\mu(g_1^+,g_2^-,g_3^+) &= A_R^{(+-+)}(p_5,p_6;g_1,g_2,g_3) = [A_R^{(+--)}(p_6,p_5;g_2,g_1,g_3)]^* \,,  \nonumber\\
 L_R^\mu(p_5^+,p_6^-) \, S_\mu(g_1^-,g_2^+,g_3^+) &= A_R^{(-++)}(p_5,p_6;g_1,g_2,g_3) = [A_R^{(+--)}(p_6,p_5;g_1,g_2,g_3)]^* \,,  \nonumber\\
%%%%%
 L_R^\mu(p_5^+,p_6^-) \, S_\mu(g_1^-,g_2^-,g_3^-) &= A_R^{(---)}(p_5,p_6;g_1,g_2,g_3) = [A_R^{(+++)}(p_6,p_5;g_1,g_2,g_3)]^*.  \label{HelAmplggg}
 \end{align}
The corresponding amplitudes for right-handed leptonic current can be obtained by simply interchanging $p_5 \leftrightarrow p_6$.
Note that the complex conjugation operation has to be applied only on the spinor structures in~\eqref{PMMg}~\eqref{PPPg},
and not on the coefficients $\alpha_j,\,\beta_j$.

The unrenormalised helicity amplitude coefficients are vectors in colour space and
have perturbative expansions:
\begin{equation}
\Omega_g^{\rm un} = 
\sqrt{4 \pi \alpha_s} \, d^{a_1 a_2 a_3}\, \left[
 \left(\frac{\alpha_s}{2\pi}\right) \Omega_g^{(1),{\rm un}}  
+ \left(\frac{\alpha_s}{2\pi}\right)^2 \Omega_g^{(2),{\rm un}} 
+ {\cal O}(\alpha_s^3) \right] \;,
\end{equation}
for $\Omega_g = \alpha_i, \beta_i$. 
The dependence on $(x,y,z)$ is again implicit.

\subsection{$gg\gamma V$: The amplitude in spinor helicity notation}
In the $gg\gamma V$-case there are three independent helicity configurations.
Two of them can be chosen identical to those in the $gggV$-case, namely $(g_1^+,g_2^-,\gamma_3^-)$ and $(g_1^+,g_2^+,\gamma_3^+)$, the third is taken as 
$(g_1^+,g_2^+,\gamma_3^-)$.

Fixing the helicities and contracting with the right-handed lepton current we have:
\begin{align}
A_R^{(+--)}&(p_5,p_6;g_1,g_2,\gamma_3) = L_R^\mu(p_5^+,p_6^-) \, S_\mu(g_1^+,g_2^-,\gamma_3^-) = 
 \frac{1}{\sqrt{2}} \, \frac{\langle 2\,3\rangle }{\langle 1\, 2\rangle \langle 1\, 3\rangle  [2\,3]} 
   \nonumber \\ \times  \Bigg\{ 
 & \langle 2\,5\rangle \langle 3\,5\rangle  [6\,5] \,\eta_1(x,y,z) 
+ \langle 2\,3\rangle \langle 2\,5\rangle  [2\,6] \, \eta_2(x,y,z)
+ \langle 2\,3\rangle \langle 3\,5\rangle  [3\,6] \, \eta_3(x,y,z) \, \Bigg\} \,,\label{PMMgamma}
\end{align}

\begin{align}
 A_R^{(+++)}&(p_5,p_6;g_1,g_2,\gamma_3) = L_R^\mu(p_5^+,p_6^-) \, S_\mu(g_1^+,g_2^+,\gamma_3^+) =
 \frac{1}{\sqrt{2}} \nonumber \\\times  \Bigg\{ 
&\frac{[1\,3] \langle 1\,5\rangle  [1\,6]}{\langle 1\, 2\rangle \langle 2\,3\rangle} \theta_1(x,y,z)
+\frac{[2\,3] \langle 2\,5\rangle  [2\,6]}{\langle 1\, 2\rangle \langle 1\,3\rangle} \theta_2(x,y,z)
+\frac{[2\,3] \langle 2\,5\rangle  [1\,6]}{\langle 1\, 2\rangle \langle 2\,3\rangle} \theta_3(x,y,z)    \Bigg\} \,,\label{PPPgamma}
\end{align}

\begin{align}
A_R^{(++-)}&(p_5,p_6;g_1,g_2,\gamma_3) = L_R^\mu(p_5^+,p_6^-) \, S_\mu(g_1^+,g_2^+,\gamma_3^-) = 
 \frac{1}{\sqrt{2}} \, \frac{[ 1\,2 ] }{\langle 1\, 2\rangle [ 1\, 3]  [2\,3]} 
 \nonumber \\ \times  \Bigg\{ 
& [ 1\,2] [ 1\,6]  \langle 1\,5 \rangle \, \tau_1(x,y,z) 
+ [ 1\,2] [ 2\,6]  \langle 2\,5 \rangle \, \tau_2(x,y,z) 
+ [ 1\,6] [ 2\,6]  \langle 6\,5 \rangle \, \tau_3(x,y,z)\, \Bigg\} \,. \label{PPMgamma}
\end{align}

From $A_R^{(+--)}(p_5,p_6;g_2,g_1,\gamma_3)$, $A_R^{(++-)}(p_5,p_6;g_1,g_2,\gamma_3)$ and $A_R^{(+++)}(p_5,p_6;g_1,g_2,\gamma_3)$
all the other helicity configurations can be obtained by parity and charge conjugation:
\begin{align}
A_R^{(-+-)}(p_5,p_6;g_1,g_2,\gamma_3) &=  A_R^{(+--)}(p_5,p_6;g_2,g_1,\gamma_3)   \nonumber \\
A_R^{(--+)}(p_5,p_6;g_1,g_2,\gamma_3) &= [A_R^{(++-)}(p_6,p_5;g_1,g_2,\gamma_3)]^* \nonumber\\
%%%%%
A_R^{(+-+)}(p_5,p_6;g_1,g_2,\gamma_3) &= [A_R^{(+--)}(p_6,p_5;g_2,g_1,\gamma_3)]^* \nonumber\\
A_R^{(-++)}(p_5,p_6;g_1,g_2,\gamma_3) &= [A_R^{(+--)}(p_6,p_5;g_1,g_2,\gamma_3)]^* \nonumber\\
%%%%%
A_R^{(---)}(p_5,p_6;g_1,g_2,\gamma_3) &= [A_R^{(+++)}(p_6,p_5;g_1,g_2,\gamma_3)]^*. 
 \end{align}
As before, the left-handed helicity amplitudes can be found by the exchange $ p_5 \leftrightarrow p_6 $,
and the complex conjugation has to be performed only on the spinor structures and not on the 
coefficients $\eta_j,\,\theta_j,\, \tau_j$.

The unrenormalised helicity amplitude coefficients are vectors in colour space and
have perturbative expansions:
\begin{equation}
\Omega_\gamma^{\rm un} = 
\sqrt{4 \pi \alpha} \, \delta^{a_1 a_2}\, \left[
 \left(\frac{\alpha_s}{2\pi}\right) \Omega_\gamma^{(1),{\rm un}}  
+ \left(\frac{\alpha_s}{2\pi}\right)^2 \Omega_\gamma^{(2),{\rm un}} 
+ {\cal O}(\alpha_s^3) \right] \;,
\end{equation}

for $\Omega_\gamma = \eta_i, \theta_i, \tau_i$. 
The dependence on $(x,y,z)$ is again implicit.

\subsection{Analytic continuation to the scattering kinematics}
\label{sec:crossings}
In order to compute the two-loop contributions to $V$-plus-jet and $V$-plus-photon production at hadron colliders, the helicity amplitudes
must be continued to the appropriate kinematical situations.

The relevant partonic subprocesses are:
\begin{eqnarray}
&&g(p_1) + g(p_2) \to g(-p_3) + V(q) \to g(-p_3) + l^+(p_5) + l^-(p_6) \;, \label{2A}\\
&&g(p_2) + g(p_3) \to g(-p_1) + V(q) \to g(-p_1) + l^+(p_5) + l^-(p_6) \;, \label{4A}
\end{eqnarray}
where the second crossing is required to fully account for all helicity combinations, 
and 
\begin{eqnarray}
&&g(p_1) + g(p_2) \to \gamma(-p_3) + V(q) \to \gamma(-p_3) + l^+(p_5) + l^-(p_6) \;. \label{2Agamma}
\end{eqnarray}

With the notation above the definitions of the helicity amplitudes in terms of momentum spinors~\eqref{PMMg}~\eqref{PPPg} and 
\eqref{PPMgamma} 
remain unchanged under crossing. Considering in fact an outgoing leptonic current defined as:
\begin{equation}
 V(q) \longrightarrow l^+(p_5) + l^-(p_6)
\end{equation}
with
\begin{equation}\label{eq_leptoniccurrentout}
L^{\mu}(p_5,p_6)\Big|_{out} = \bar{u}(p_6) \, \gamma^{\mu} \, v(p_5),
\end{equation}
we find that:
\begin{align*}
 L_R^\mu(p_5^+,p_6^-)\Big|_{in} &= [ 6 \, | \gamma^\mu | 5 \rangle = L_R^\mu(p_5^-,p_6^+)\Big|_{out} \\
 L_L^\mu(p_5^-,p_6^+)\Big|_{in} &= [ 5 \, | \gamma^\mu | 6 \rangle = L_L^\mu(p_5^+,p_6^-)\Big|_{out}.
\end{align*}
This means that the expressions for the helicity amplitudes defined in the two sections above
remain unchanged provided that $p_5$ is now considered as the label of the antilepton and $p_6$ the one of the lepton.

Special care has to be taken in the analytic continuation 
of the helicity coefficients $\Omega_g$ and $\Omega_\gamma$.
In the kinematical situation in~\eqref{2A} and~\eqref{2Agamma}
$q^2$ remains time-like, but only $s_{12}$ becomes positive:
\begin{equation}
q^2 > 0\;,\quad s_{12} > 0\;,\quad s_{13} < 0, \quad s_{23} < 0\; ,
\end{equation}
or, equivalently,
\begin{equation}
x>0\;, \quad y<0 \;, \quad z<0\; .
\end{equation}
As shown in~\cite{ancont}  (where this region is denoted as (2a)$_+$) 
and used for example in~\cite{Vgamma}, 
this kinematical situation can be expressed by introducing new dimensionless variables
\begin{equation} 
u_1 = -\frac{s_{13}}{s_{12}}=-\frac{y}{x}\,, \qquad v_1 = \frac{q^2}{s_{12}} =
 \frac{1}{x}\;,
\end{equation}
which fulfil
\begin{displaymath}
0\leq u_1 \leq 1-v_1\,, \qquad 0\leq v_1 \leq 1\;.
\end{displaymath}

To account for all helicity combinations in the case of $gg \to gV$, also the kinematical situation~\eqref{4A} must be considered.
In this case we have
\begin{equation}
q^2 > 0\;,\quad s_{12} < 0\;,\quad s_{13} < 0, \quad s_{23} > 0\; ,
\end{equation}
This can be treated with the following choice of variables~\cite{ancont} (this region is denoted as
(4a)$_+$) :
\begin{equation} 
u_2 = -\frac{s_{13}}{s_{23}}=-\frac{y}{z}\,, \qquad v_2 = \frac{q^2}{s_{23}} =
 \frac{1}{z}\;,
\end{equation}
which fulfil again
\begin{displaymath}
0\leq u_2 \leq 1-v_2\,, \qquad 0\leq v_2 \leq 1\;.
\end{displaymath}

Note that the two kinematical regions~\eqref{2A} and~\eqref{4A} are turned each other 
 by the permutation $p_1 \leftrightarrow p_3$,
in particular one has:
\begin{align*}
 u_1(p_1 \leftrightarrow p_3) &=  u_2\\
 v_1(p_1 \leftrightarrow p_3) &=  v_2.
\end{align*}
As shown in~\eqref{HelAmplggg}, in the $gggV$-case, in order to obtain all the different helicity configurations, 
we also need to exploit the Bose symmetry of the external gluons.
It is now clear that whenever the permutation $p_1 \leftrightarrow p_3$ is performed, 
this only amounts to switching from region~\eqref{2A} to region~\eqref{4A}. 

We provide the one-loop and two-loop coefficients in all relevant regions in Mathematica format 
together with the arXiv-submission of this paper.

\section{Outline of the calculation}
\label{sec:calc}
The two-loop corrections to the coefficients $\Omega_b$ 
can be evaluated through a calculation of the relevant Feynman diagrams.
The calculation proceeds as follows.
The diagrams contributing to the process are produced 
using QGRAF~\cite{qgraf}. 
In the $gggV$-case there are $12$ diagrams at one loop and $264$ at two loops,
while in the $gg\gamma V$-case there are $8$ diagrams at one and $138$ at two loops.
The tensor coefficients are evaluated analytically diagram by diagram 
applying the projectors defined above. 
As a result, one obtains the tensor coefficients in terms of thousands of 
planar and non-planar two-loop scalar integrals, which can be
classified in two auxiliary topologies, one planar and the other non-planar~\cite{3jme}.
In order to do so, one needs to perform both shifts in the integration variables and
permutations on the external legs. All the routines needed for this purpose have been coded in FORM~\cite{form}
and checked against the new automated shift-finder implemented in Reduze2~\cite{reduze2}.
Through the usual IBP identities~\cite{chet1} one can reduce independently all the integrals
belonging to these two auxiliary topologies to a small set
of master integrals. This reduction is performed using the Laporta algorithm~\cite{laporta}
implemented in the Reduze code~\cite{reduze2,reduze}.
All the masters for the topologies above are known as series in
the parameter $\epsilon = (4-d)/2$ through a systematic approach based
on the differential equation method~\cite{de2,3jmaster}. The masters are 
expressed as Laurent expansion in $\e$, with coefficients containing 
harmonic polylogarithms (HPLs,~\cite{hpl}) and two-dimensional harmonic 
polylogarithms (2dHPLs,~\cite{3jmaster}). Numerical implementations of these 
functions are available~\cite{hplnum}. 
For all the intermediate algebraic manipulations we have made extensive use of 
FORM~\cite{form} and Mathematica~\cite{mathematica}.
The two-loop unrenormalised helicity coefficients $\Omega_b^{(2),{\rm un}}$
can then be evaluated as linear combination of the tensor coefficients.
The whole computation is performed in the euclidean non-physical region, where
the amplitude is real. The final result is then analytically continued
to the physical regions relevant for $Z+jet/\gamma$ production at LHC, as thoroughly
discussed in~\cite{ancont} and in section~\ref{sec:crossings}.

\subsection{UV Renormalisation and IR subtraction}
\label{sec:ir}
Renormalisation of ultraviolet divergences is 
performed in the \MSbar\ scheme by replacing 
the bare coupling $\alpha_0$ with the renormalised coupling 
$\alpha_s\equiv \alpha_s(\mu^2)$,
evaluated at the renormalisation scale $\mu^2$. 
Since there is no tree level contribution to the amplitude,
we only need the one loop relation between the bare and renormalised couplings:
\begin{equation}
\alpha_0\mu_0^{2\e} S_\e = \alpha_s \mu^{2\e}\left[
1- \frac{\beta_0}{\e}\left(\frac{\alpha_s}{2\pi}\right) 
+{\cal O}(\alpha_s^2) \right]\; ,
\end{equation}
where
\begin{displaymath}
S_\e =(4\pi)^\e e^{-\e\gamma}\qquad \mbox{with Euler constant }
\gamma = 0.5772\ldots
\end{displaymath}
and $\mu_0^2$ is the mass parameter introduced 
in dimensional regularisation to maintain a 
dimensionless coupling 
in the bare QCD Lagrangian density.
$\beta_0$ is the first 
coefficient of the QCD $\beta$-function:
\begin{equation}
\beta_0 = \frac{11 \CA - 4 T_R \NF}{6},
\end{equation}
with the QCD colour factors
\begin{equation}
\CA = N,\qquad C_F = \frac{N^2-1}{2N},
\qquad T_R = \frac{1}{2}\; .
\end{equation}
The renormalisation is performed at fixed scale $\mu^2 = q^2$. The renormalised 
helicity coefficients read:
\begin{eqnarray}
\Omega_g^{(1)}  &=& 
S_\e^{-1} \Omega_g^{(1),{\rm un}} ,  \nonumber \\
\Omega_g^{(2)} &=& 
S_\e^{-2} \Omega_g^{(2),{\rm un}}  
-\frac{3 \beta_0 }{ 2 \epsilon} S_\e^{-1}
\Omega_g^{(1),{\rm un}}\;.
\end{eqnarray}

\begin{eqnarray}
\Omega_\gamma^{(1)}  &=& 
S_\e^{-1} \Omega_\gamma^{(1),{\rm un}} ,  \nonumber \\
\Omega_\gamma^{(2)} &=& 
S_\e^{-2} \Omega_\gamma^{(2),{\rm un}}  
-\frac{\beta_0 }{\epsilon} S_\e^{-1}
\Omega_\gamma^{(1),{\rm un}}\;.
\end{eqnarray}

After performing ultraviolet renormalisation, the amplitudes still
contain singularities, which are of infrared origin and will be  analytically
cancelled by those occurring in radiative processes of the
same order.
Catani~\cite{catani} has shown how to organise the 
infrared pole structure of the one- and two-loop contributions renormalised in the 
$\overline{{\rm MS}}$-scheme in terms of the tree and renormalised one-loop amplitudes.
The same procedure applies to the tensor coefficients. 
Since there is no tree level
process contributing, their pole structure can be 
separated off as follows:
\begin{eqnarray}
\Omega_{b}^{(1)} &=& \Omega_{b}^{(1),{\rm finite}},\nonumber \\
\Omega_{b}^{(2)} &=& {\bom I}_{b}^{(1)}(\epsilon) \Omega_{b}^{(1)}+ \Omega_{b}^{(2),{\rm finite}},
\end{eqnarray}
where again $b = g,\gamma$.

In the two cases the operator
$\bom{I}_{b}^{(1)}(\epsilon)$ 
is given by
\begin{align}
\bom{I}_g^{(1)}(\epsilon) &
= - N \, \frac{e^{\epsilon\gamma}}{2\Gamma(1-\epsilon)} \Biggl[
\left(\frac{1}{\epsilon^2}+\frac{\beta_0}{N\,\epsilon} \right)
( {\tt S}_{12}+{\tt S}_{13}+{\tt S}_{23})\Biggr ]\; ,\\
\bom{I}_\gamma^{(1)}(\epsilon) &
= - N \, \frac{e^{\epsilon\gamma}}{\Gamma(1-\epsilon)} \Biggl[
\left(\frac{1}{\epsilon^2}+\frac{\beta_0}{N\,\epsilon} \right)\,
{\tt S}_{12}\Biggr ]\; ,\label{eq:I1}
\end{align}
where, since we have set $\mu^2 = s_{123}$:
\begin{align}
{\tt S}_{ij} &= \left(-\frac{s_{123}}{s_{ij}}\right)^{\epsilon}
\end{align}
Note that on expanding ${\tt S}_{ij}$,
imaginary parts are generated, depending on which kinematical configuration we are working in.
In the decay kinematics $Z \to ggg\, / \, gg\gamma$ for example we have that all the $s_{ij}$ become positive,
so that all three terms will generate imaginary parts whose sign is fixed by the small imaginary
part $+i0$ of $s_{ij}$.
On the other hand if we are interested in the scattering kinematics $gg \to Zg \,/\, Z\gamma$ 
only $s_{12}$ or $s_{13}$ become positive, with the usual $s_{ij}+i 0$ prescription.

For the infrared factorisation of the two-loop results, 
the renormalised one-loop helicity amplitude coefficients
are needed through to ${\cal O}(\epsilon^2)$.
Their decomposition in colour structures is straightforward, namely the whole colour dependence is 
in the overall factors $d^{a_1 a_2 a_3}$ and $\delta^{a_1 a_2}$ for $gggV$ and $gg\gamma V$ respectively.
\begin{equation}
\Omega^{(1),{\rm finite}}_b(x,y,z) = \, a_{\Omega_b}(x,y,z) \;. 
\label{eq:oneloopamp}
\end{equation}
The expansion of the coefficients through to $\e^2$ yields HPLs and 2dHPLs 
up to weight 4. The explicit expressions are of considerable size, 
such that we only quote the $\e^0$-terms in the appendix. To this order, 
the coefficients had been derived previously~\cite{glover} in terms of 
logarithms and dilogarithms. 
The expressions through to ${\cal O}(\e^2)$
in Mathematica format are appended to the arXiv submission 
of this article.

The finite two-loop remainder is obtained by subtracting the
predicted infrared structure (expanded through to ${\cal O}(\epsilon^0)$) from
the renormalised helicity coefficient.  We further decompose it according to the colour structures as follows: 
\begin{align}
\Omega^{(2),{\rm finite}}_b(x,y,z) &= N \, A_{\Omega_b} + {1 \over N}\, B_{\Omega_b} + N_f \, C_{\Omega_b}. \; , 
\label{eq:twoloopamp}
\end{align}

The helicity coefficients contain HPLs and 2dHPLs up to weight 4.
The size of each helicity coefficient is comparable 
to the size of the helicity-averaged tree times two-loop matrix element 
for $3j$ production quoted in~\cite{3jme}, and we decided not to
include them here explicitly. The complete set of coefficients in Mathematica 
format is attached to the arXiv submission of this article.

\subsection{Simplification using the Symbol formalism}
After the computation of the amplitudes and subtraction of UV- and IR-divergences we used an in-house implementation of the algorithm described in \cite{Duhr:2011zq} to express the result as far as possible in logarithms and polylogarithms of functions of the kinematic invariants. The GiNaC libary was used to evaluate the 2dHPLs \cite{Vollinga:2005pk} and the implementation of the PSLQ algorithm contained in the arprec library \cite{arprec} to find the parts mapped to zero by the symbol map. 

It is well known that up to transcendental weight three all two-dimensional harmonic polylogarithms can be expressed this way. However, for weight four this is not always the case. In \cite{Goncharov_delta} it was conjectured that a combination of 2dHPLs can be expressed in logarithms and polylogarithms if and only if its symbol fulfills a certain symmetry condition. In the present case, we found this condition in general not to be fulfilled and were also not able to express our result in logarithms and polylogarithms only. Nevertheless we reduced the number of required functions in all kinematic regions as far as possible, having to resort to 17 2dHPLs of weight four.

In the past, surprising relations between certain QCD and N=4 SYM amplitudes have been found, for example in the case of $H\rightarrow ggg$ at two loops in the 
heavy-top-limit \cite{Hggg,Brandhuber:2012vm,Duhr:2012fh}. In the leading color part of the finite 
two-loop amplitude, the weight four contribution without a rational factor was 
found~\cite{Brandhuber:2012vm} to be 
helicity-independent and
equal to the three-point form factor remainder function in planar N=4 SYM. 
In the present cases, however, no such relation could be observed. This feature can be understood 
from the fact that, in contrast to the Higgs amplitudes, no purely gluonic contribution is 
present here, due to the internal quark loop  coupling to the vector boson.

\section{Checks on the result}
\label{sec:checks}
Several non-trivial checks were applied to validate our results.
\begin{enumerate}
 \item As a first check we computed all 14 tensor coefficients in  (\ref{eq:smubase}) at one-loop
       order for the $gggV$-case, and we verified that we can reproduce the results in~\cite{glover} up to order $O(\epsilon^0)$.
       Performing this check was not entirely trivial. In~\cite{glover} the results for the one-loop
       helicity amplitudes are given in the case of an on-shell $Z$ with a fixed polarization.
       Moreover, the amplitudes for different helicity configurations are given choosing an explicit representation
       for the polarization vectors of the external particles. This representation does not respect the gauge 
       choice performed in~\eqref{Gauge}, so that we cannot naively start from our tensor structure and 
       fix the polarization vectors in the same way to reproduce their result.
       Nevertheless, as explained in section~\ref{sec:tensor}, the full gauge-invariant tensor can be fully reconstructed
       taking suitable linear combinations of the tensor coefficients of the gauge-fixed tensor.
       Once the gauge-invariant tensor is known, one can then use the explicit representation of the polarization
       vectors given in~\cite{glover} and 
       demonstrate the analytic agreement of the expressions.

 \item We computed all the 14 tensor coefficients both at one-loop and at two-loop order, in the $gggV$- 
       and in the $gg\gamma V$-case.
       Following the procedure outlined in section~\ref{sec:tensor}, 
       we obtained the 14 coefficients of the gauge invariant tensor
       for both processes, and we verified that they respect the expected symmetry relations under 
       permutation of the external gluons.

 \item The IR singularity structure of our results agrees with the prediction 
       of Catani formula~\cite{catani}, see section~\ref{sec:ir}.

 \item We compared the helicity amplitudes $\Omega_b^{(1)}$ for the $gggV$- and the $gg\gamma V$-case.
       We verified the following identities for the one-loop amplitude coefficients:
       \begin{align}
        2\, a_{\alpha_j}(x,y,z) &= a_{\eta_j}(x,y,z), \nonumber \\
        2\, a_{\beta_j}(x,y,z) &= a_{\theta_j}(x,y,z). \qquad j=1,2,3 \,.
       \end{align}

 \item Finally, we performed the same comparison at two-loop order, finding:
        \begin{align}
        2\, B_{\alpha_j}(x,y,z) &= B_{\eta_j}(x,y,z), \nonumber \\
        2\, B_{\beta_j}(x,y,z) &= B_{\theta_j}(x,y,z), \qquad j=1,2,3\,,
       \end{align}
       which follow from the structure of the underlying two-loop diagrams. The subleading 
       colour coefficients $B$ are unaffected by renormalisation and infrared subtraction. 
      No relation of this type can be found for the coefficients $C_{\Omega_b}$, which are 
      determined purely from renormalisation counterterms and IR subtraction,
      which differ in the cases $b=g,\gamma$.
\end{enumerate}

\section{Conclusions and Outlook}
\label{sec:conc}
In this paper we presented the two-loop corrections to the helicity amplitudes for the processes $gg \to V g$ and $gg \to V \gamma$.
We performed the calculation in dimensional regularisation by applying $d$-dimensional projection operators to the most general
tensor structure of the amplitude. 
We showed how an explicit gauge choice can
reduce considerably the complexity of the basic tensor structures appearing while retaining the 
full information on the gauge-invariant amplitudes. 
We expressed our results in terms of dimensionless 
helicity coefficients, which multiply four-dimensional spinor structures.
We extracted the infrared singularities by means of an infrared factorisation formula and 
provide compact analytic expressions for the finite part of the two-loop helicity coefficients in all relevant kinematical regions.

The matrix elements derived here contribute to the NLO corrections to the gluon-induced 
production of $Z\gamma$ and $Z+j$ final states at the LHC. Viewed in 
an expansion in the strong coupling constant, these  contributions are 
 formally N$^3$LO as far as the 
reactions $p p \to V \gamma+ X $, $p p \to V j+ X$ are concerned. However, due 
 to the large gluon-gluon luminosity at the LHC, these contributions 
could be comparable in size with the 
NNLO corrections to $q \bar{q} \to V g$, $q g \to V q$ and  $q \bar{q} \to V \gamma$. Their 
inclusion will also help to stabilise the substantial scale dependence of the gluon-induced 
subprocesses, which were known only at Born-level up to now.

\section*{Acknowledgements}
We are grateful to C.\ Duhr for many interesting discussions and for 
some comments on the manuscript. LT wishes to thank A.\ von Manteuffel for his kind assistance in the use of Reduze2.

This research was supported in part by
the Swiss National Science Foundation (SNF) under contract
PDFMP2-135101 and  200020-138206, as well as  by the European Commission through the 
``LHCPhenoNet" Initial Training Network PITN-GA-2010-264564.  

\appendix
\section{One-loop helicity amplitudes}
\label{app:1loopDecay}
\subsection{$V \to ggg $ at one loop}
We reproduce the leading order $O(\epsilon^2)$ for $V \to ggg $. The complete expressions up to order $O(\epsilon^2)$ can be found in the attachments of the arXiv version of this paper.

\begin{align}
a_{\alpha_{1}}(x,y,z) &= 2 x  \left(\frac{1}{1-x}-\frac{2}{z}\right) \;\log(x) + 2 y \left(\frac{1}{1-y}-\frac{2}{z}\right) \; \log(y)  \nonumber \\ 
&-2  \left(\frac{(1-x)x+(1-y) y}{z^2}\right) \Bigg[ \frac{\pi ^2}{6} + \log(x) \log(y)  \nonumber \\
&   \qquad \qquad \qquad \qquad \qquad \qquad -(\, \log(1-x) \log(x)+\text{Li}_{2}(x) \,) \nonumber \\
&   \qquad \qquad \qquad \qquad \qquad \qquad -(\, \log(1-y) \log(y)+\text{Li}_{2}(y) \,) \; \vphantom{\frac12} \Bigg] \,,
\end{align}

\begin{align}
a_{\alpha_{2}}(x,y,z) &= 2 y  \left(\frac{1}{1-y}-\frac{1}{z}\right) - \frac{2 x(2 y+z)}{z^2} \log(x) 
- \frac{2 x y (z+(1-y)(2 y+z))}{(1-y)^2 z^2} \log(y) \nonumber  \\
&+2 \left(-\frac{x \left(2 y^2+2 y z+z^2\right)}{z^3}\right) \Bigg[ \frac{\pi ^2}{6}  + \log(x) \log(y)  \nonumber  \\
&   \qquad \qquad \qquad \qquad \qquad \qquad -(\, \log(1-x) \log(x)+\text{Li}_{2}(x) \,) \nonumber  \\
&   \qquad \qquad \qquad \qquad \qquad \qquad -(\, \log(1-y) \log(y)+\text{Li}_{2}(y) \,) \; \vphantom{\frac12} \Bigg]\,, 
\end{align}

\begin{align}
a_{\alpha_{3}}(x,y,z) = - a_{\alpha_{2}}(y,x,z)\,,
\end{align}
\begin{align}
a_{\beta_{1}}(x,y,z) =  - 2 \left(1 - \frac{1}{y}\right)\,,\quad &  a_{\beta_{2}}(x,y,z) =  - 2 \left(1 - \frac{1}{z}\right)\,, \nonumber \\
a_{\beta_{3}}(x,y,z) = & - 4 \,.
\end{align}

\subsection{$V \to gg\gamma $ at one loop}
At one loop, the two amplitudes are related to each other as follows:
\begin{align}
a_{\eta_i}(x,y,z) &=  2 \,a_{\alpha_i}(x,y,z) \quad \text{for}\, i = 1,2,3 \,, \nonumber \\ 
a_{\theta_i}(x,y,z)& =  2 \,a_{\beta_i}(x,y,z) \quad \text{for}\, i = 1,2,3 \,, 
\end{align}
\begin{align}
a_{\tau_1}(x,y,z) = 2 \,a_{\alpha_3}(z,y,x) &\,, \quad
a_{\tau_2}(x,y,z) = 2 \,a_{\alpha_2}(z,y,x) \,, \nonumber \\ \qquad a_{\tau_3}(x,y,z) &=  2 \,a_{\alpha_1}(z,y,x) \,
\end{align}

\section{Two-loop amplitudes: all-plus helicity coefficients}
\label{app:2loopDecay}
Due to the length of the resulting expressions we chose to reproduce only the all-plus $(g_1^+,g_2^+,g_3^+/\gamma_3^+)$ helicity amplitudes of both processes in the decay region, which are considerably shorter than the 
other helicity combinations
 and contain only functions up to transcendental weight two. The full result can be found in the attachments to the arXiv submission of this paper in Mathematica format.
\subsection{$V \to ggg $ at two loops }
The coefficients for the $(g_1^+,g_2^+,g_3^+)$ helicity configuration are:
\begin{align}
 A_{\beta_1}(x,y,z) = & -\frac{1}{27} \left(27 \left(3-\frac{1}{1-x}-\frac{1}{y}-\frac{1}{1-z}\right)-\frac{4 z}{x}-\frac{4 z^2}{x^2} \right. \nonumber\\
& \left. \quad \qquad +x^2  \left(-\frac{4}{z^2}-\frac{4z}{y^3}\right)+\frac{4 x}{z} \left(-1-\frac{z^3}{y^3}\right)\right) -\frac{11}{2} \left(1-\frac{1}{y}\right)  i \pi  \nonumber \\ 
&-\frac{1}{12}  \left(\frac{3 (1-y)}{x y}-\frac{2 (-1+2 y)}{y^2}+\frac{3 (1-y)}{y z}+\frac{14 (1-y) z}{y^3}-\frac{14 z^2}{y^3}\right)  \pi ^2 \nonumber \\ 
&+\frac{1}{6} \left(11+\frac{6}{(1-x)^2}-\frac{6}{1-x}-\frac{42 x+11 y}{y^2}\right)  \log(x) + \frac{1}{6} \left(11-\frac{47}{y}\right)  \log(y) \nonumber \\ 
&+\frac{1}{6}  \left(11+\frac{-42+42 x+31 y}{y^2}+\frac{6}{(1-z)^2}-\frac{6}{1-z}\right) \log(z) \nonumber \\ 
&-\frac{1}{2}  \left(\frac{2}{y}+\frac{3 x}{y z}\right)  \log(x) \log(y) -\frac{1}{2} \left(\frac{2}{y}+\frac{3 z}{x y}\right)  \log(y) \log(z) \nonumber \\ 
&-  \left(\frac{7 (1-x) x}{y^3}+\frac{1}{y^2}-\frac{7 x}{y^2}-\frac{1}{y}\right)  \log(x) \log(z) \nonumber \\ 
&+\frac{1}{2}  \left(\frac{2 (1-7 z)}{y^2}-\frac{3}{z}+\frac{3 (1-z)}{yz}+\frac{14 (1-z) z}{y^3}\right)  \nonumber \\
& \qquad \qquad \qquad \qquad  \qquad \qquad  \qquad \qquad \times \left( \log(1-x) \log(x)+\text{Li}_{2}(x)\right)  \nonumber \\ 
&+\frac{1}{2}  \left(\frac{4}{y}+\frac{3 x}{y z}+\frac{3 z}{xy}\right) \left( \log(1-y) \log(y)+\text{Li}_{2}(y)\right)  \nonumber \\ 
&-\frac{1}{2}  \left(\frac{14 x^2}{y^3}-\frac{14 x (1-y)}{y^3}-\frac{3 (1-y)}{x y}+\frac{-2+3y}{y^2}\right)  \nonumber \\
& \qquad \qquad \qquad \qquad  \qquad  \qquad  \qquad \qquad \times \left( \log(1-z) \log(z)+\text{Li}_{2}(z)\right)\,,   
\end{align}

\begin{align}
 A_{\beta_2}(x,y,z) = &-\frac{1}{27}  \left(\frac{27}{1-x}+\frac{27}{1-y}-\frac{4 y}{x}
-\frac{4 y^2}{x^2}+x^2 \left(-\frac{4}{y^2}-\frac{4 y}{z^3}\right) \right.\nonumber  \\
& \quad \left. + \frac{4 x}{y} \left(-1-\frac{y^3}{z^3}\right)-\frac{54}{z}\right) - \frac{11}{2} \left(1-\frac{1}{z}\right)  i \pi \nonumber\\
& -\frac{1}{12}  \left(\frac{x^2 (14+y (-58+45y))}{y^2 z^2} +\frac{14x^3 (-2+3 y)}{y^2 z^2}-\frac{2 (1-y) y (-8+21 y)}{x z^2} \right. \nonumber \\
& \left. -\frac{4 x (1-y) (-4+5 y)}{y z^2}+\frac{3 (2+3 y (-4+5 y))}{z^2}+\frac{14 (1-y)^2 y^2}{x^2 z^2}+\frac{14 x^4}{y^2 z^2}\right)  \pi ^2 \nonumber \\ 
&-\frac{1}{6}  \left(\frac{53 x}{z}+\frac{42 x^2}{y z}-\frac{(-47-53 (-2+x) x) y}{(1-x)^2 z}\right) \log(x) \nonumber \\ 
&+\frac{1}{6}  \left(-\frac{53 y}{z}-\frac{42 y^2}{x z}-\frac{x (47+53 (-2+y) y)}{(1-y)^2 z}\right)  \log(y) \nonumber \\ 
&-\frac{1}{6}  \left(31+\frac{42 x}{y}+\frac{42y}{x}+\frac{11}{z}\right)  \log(z) \nonumber \\ 
&-\frac{1}{2} \left(\frac{x^2}{z^2}+\frac{x (2-12 y)}{z^2}+\frac{y (2+y)}{z^2}\right)  \log(x) \log(y) \nonumber \\ 
&-\frac{1}{2}  \left(1+\frac{14 (1-z)^2}{y^2}-\frac{2 (1-z) (-1+7 z)}{y z}\right)  \log(x)\log(z) \nonumber \\ 
&-\frac{1}{2}  \left(1+\frac{14 y^2}{x^2}+\frac{-2+14 y+\frac{2}{z}}{x}\right)  \log(y) \log(z) \nonumber \\ 
&+\frac{1}{2}  \left(\frac{14(1-z)^2}{y^2}+\frac{14 y^2}{z^2}-\frac{14 y (1-z)}{z^2}-\frac{2 (1-z) (-1+7 z)}{y z} \right.\nonumber \\
& \qquad\qquad\qquad\qquad \qquad  \left. +\frac{3+2 (-2+z) z}{z^2}\right) \left( \log(1-x) \log(x)+\text{Li}_{2}(x)\right)  \nonumber \\ 
&+\frac{1}{2} \left(\frac{14 (1-z)^2}{x^2}+\frac{14 x^2}{z^2}-\frac{14 x (1-z)}{z^2}-\frac{2 (1-z) (-1+7 z)}{x z} \right.\nonumber \\
& \qquad\qquad \qquad\qquad\qquad \left. +\frac{3+2 (-2+z) z}{z^2}\right) \left( \log(1-y)\log(y)+\text{Li}_{2}(y)\right)  \nonumber \\ 
&- \left(\frac{1-x}{x}-\frac{7 x^2}{y^2}+\frac{x (-8+7 x)}{(1-x) y}-\frac{7 y}{x}-\frac{7 y^2}{x^2}-\frac{1}{(1-x)x z}\right)  \nonumber \\
& \qquad \qquad \qquad \qquad\qquad \qquad \qquad \qquad \qquad \times \left( \log(1-z) \log(z)+\text{Li}_{2}(z)\right) \,, 
\end{align}

\begin{align}
 A_{\beta_3}(x,y,z) =   & -\frac{1}{27} \left(81-\frac{8 y^2}{z^2}-\frac{8 z}{y}+\frac{8 (1-z) z}{y^2}
+\frac{8 y}{z^2} \left(1-z+\frac{z^3}{x^2}\right)\right)  - 11 i \pi  \nonumber\\ 
 &-\frac{1}{12}  \left(3-\frac{5}{x}-\frac{14 (1-x) x}{y^2}+\frac{-5+14 x}{y}-\frac{14 (1-x) y}{x^2} \right. \nonumber \\
& \qquad \left. +\frac{14 y^2}{x^2}-\frac{14 (1-x)x}{z^2}-\frac{5-14 x}{z}\right)  \pi ^2 \nonumber \\ 
 &+\frac{1}{3}  \left(32-\frac{3}{1-x}+\frac{21 (1-x) x}{y z}\right)  \log(x) \nonumber \\ 
 &+\frac{1}{3} \left(32-\frac{3}{1-y}+\frac{21 (1-y) y}{x z}\right)  \log(y) \nonumber \\ 
 &+\frac{1}{3}  \left(32-\frac{3}{1-z}+\frac{21 (1-z) z}{x y}\right)  \log(z) \nonumber \\ 
 &-\frac{1}{2}  \left(\frac{14 x^2}{z^2}-\frac{14 x (1-z)}{z^2}+\frac{-5+z}{z}\right) \log(x) \log(y) \nonumber \\ 
 &-\frac{1}{2}  \left(1-\frac{14(1-z) z}{y^2}+\frac{-5+14 z}{y}\right)  \log(x) \log(z) \nonumber \\ 
 &-\frac{1}{2}  \left(1-\frac{14 (1-y) y}{x^2}+\frac{-5+14 y}{x}\right) \log(y) \log(z) \nonumber \\ 
 &+\frac{1}{2}  \left(2-\frac{(1-x)}{y^2 z^2} \left(14 (1-x)^2 x-(1-x) (-5+42 x) y+(-5+42 x) y^2\right)\right) \nonumber \\
& \qquad \qquad \qquad \qquad  \qquad \qquad  \qquad \qquad\qquad\times\left( \log(1-x) \log(x)+\text{Li}_{2}(x)\right)  \nonumber \\ 
 &+\frac{1}{2}  \left(2-\frac{(1-y)}{x^2 z^2} \left(14 (1-y)^2 y-(1-y) (-5+42 y) z+(-5+42 y) z^2\right)\right)  \nonumber \\
& \qquad \qquad \qquad \qquad  \qquad \qquad  \qquad \qquad\qquad\times\left( \log(1-y) \log(y)+\text{Li}_{2}(y)\right)  \nonumber \\ 
 &+\frac{1}{2}  \left(\frac{14 x^2}{y^2}-\frac{14 x (1-y)}{y^2}-\frac{14 (1-y) y}{x^2}+\frac{-5+2y}{y}+\frac{-5+14 y}{x}\right)\nonumber \\
& \qquad \qquad \qquad \qquad  \qquad \qquad  \qquad \qquad \qquad\times \left( \log(1-z) \log(z)+\text{Li}_{2}(z)\right) \,, 
\end{align}

\begin{align}
 B_{\beta_1}(x,y,z) = & \frac{1}{1-x}+\frac{1}{y}+\frac{1}{1-z}-3 \nonumber  \\ 
&+\frac{1}{12}  \left(\frac{2}{y}-\frac{1-y}{x y}-\frac{1-y}{y z}-\frac{2 (1-y)z}{y^3}+\frac{2 z^2}{y^3}\right)  \pi ^2 \nonumber \\ 
&+ \left(\frac{x}{(1-x)^2}-\frac{x}{y^2}\right)  \log(x) - \left(\frac{z}{y^2}-\frac{z}{(1-z)^2}\right)  \log(z) \nonumber \\ 
&-\frac{1}{2} \left(\frac{x}{y z}\right)  \log(x) \log(y) - \left(\frac{xz}{y^3}\right)  \log(x) \log(z) \nonumber \\ 
&-\frac{1}{2}  \left(\frac{z}{x y}\right)  \log(y) \log(z) \nonumber \\ 
&+\frac{1}{2} \left(\frac{x}{y z}+\frac{2 x z}{y^3}\right) \left( \log(1-x) \log(x)+\text{Li}_{2}(x)\right)  \nonumber \\ 
&+\frac{1}{2}  \left(\frac{x}{y z}+\frac{z}{x y}\right)\left( \log(1-y) \log(y)+\text{Li}_{2}(y)\right)  \nonumber \\ 
&-\frac{1}{2}  \left(-\frac{2 x z}{y^3}-\frac{z}{x y}\right) \left( \log(1-z) \log(z)+\text{Li}_{2}(z)\right)\,, 
\end{align}

\begin{align}
 B_{\beta_2}(x,y,z) = & -\frac{1}{1-x}-\frac{1}{1-y}+\frac{2}{z}  \nonumber\\ 
&- \frac{1}{12}  \left(3+\frac{2 x^2}{y^2}+\frac{2 x}{y}+\frac{2 y}{x}+\frac{2y^2}{x^2}
+\frac{1-2 (1-x) x}{z^2}-\frac{2 (1-x)}{z}\right)  \pi ^2 \nonumber \\ 
&- \left(\frac{2 x}{(1-x)^2}-\frac{x}{z}+\frac{x^2 z}{(1-x)^2y}\right)  \log(x) - \left(1+\frac{x}{y}+\frac{y}{x}\right)  \log(z) \nonumber \\ 
&+ \left(1-\frac{1}{(1-y)^2}+y \left(-\frac{y}{x (1-y)}+\frac{1}{z}\right)\right)  \log(y) \nonumber \\ 
&-\frac{1}{2} \left(\frac{x^2}{z^2}+\frac{y^2}{z^2}\right)  \log(x) \log(y) -\frac{1}{2} \left(1+\frac{2 x^2}{y^2}+\frac{2 x}{y}\right)  \log(x) \log(z) \nonumber \\ 
&-\frac{1}{2}  \left(1+\frac{2 y}{x}+\frac{2 y^2}{x^2}\right) \log(y) \log(z) \nonumber \\ 
&+\frac{1}{2}  \left(2+\frac{2 x (1-z)}{y^2}+\frac{1-2 (1-x) x}{z^2}-\frac{2 (1-x)}{z}\right) \nonumber \\
& \qquad \qquad \qquad \qquad \qquad \qquad \times\left( \log(1-x)\log(x)+\text{Li}_{2}(x)\right)  \nonumber \\ 
&+\frac{1}{2}  \left(2+\frac{2 y}{x}+\frac{2 y^2}{x^2}+\frac{1-2 (1-x) x}{z^2}-\frac{2 (1-x)}{z}\right)  \nonumber \\
& \qquad \qquad \qquad \qquad  \qquad \qquad \times\left( \log(1-y)\log(y)+\text{Li}_{2}(y)\right)  \nonumber \\ 
&+ \left(1+\frac{x^4+x^3 y+x y^3+y^4}{x^2 y^2}\right) \left( \log(1-z) \log(z)+\text{Li}_{2}(z)\right) \,, 
\end{align}

\begin{align} 
 B_{\beta_3}(x,y,z) =   & -3  -\frac{1}{12}  \left(3-\frac{1}{x}-\frac{2 (1-x) x}{y^2}+\frac{-1+2 x}{y}-\frac{2 (1-x) y}{x^2} \right. \nonumber  \\
& \qquad \qquad \left. +\frac{2 y^2}{x^2}-\frac{2 (1-x)x}{z^2}-\frac{1-2 x}{z}\right)  \pi ^2 \nonumber \\ 
&- \left(\frac{x}{1-x}-\frac{x}{y}-\frac{x}{z}\right)  \log(x)  + \left(\frac{y}{x}+\frac{xy}{(1-y) z}\right)  \log(y) \nonumber \\ 
&+ \left(\frac{1-x}{y}-\frac{1}{1-z}+\frac{z}{x}\right) \log(z) \nonumber \\ 
&+\frac{1}{2}  \left(\frac{(1-x) x}{z^2}+\frac{(1-y) y}{z^2}\right)  \log(x) \log(y) \nonumber \\ 
&-\frac{1}{2}  \left(1-\frac{2 (1-z) z}{y^2}+\frac{-1+2 z}{y}\right)  \log(x) \log(z) \nonumber \\ 
&-\frac{1}{2}  \left(1-\frac{2(1-y) y}{x^2}+\frac{-1+2 y}{x}\right)  \log(y) \log(z) \nonumber \\ 
&+\frac{1}{2}  \left(2- \frac{1}{y^2 z^2}\left((1-x) \left(2 (1-x)^2 x+y+x (-7+6 x) y+(-1+6x) y^2\right) \right) \right)  \nonumber \\
& \qquad \qquad \qquad \qquad  \qquad \qquad\qquad \qquad\qquad \qquad \times\left( \log(1-x) \log(x)+\text{Li}_{2}(x)\right)  \nonumber \\ 
& +\frac{1}{2}  \left(2-\frac{1}{x^2 z^2}\left((1-y) \left(2 (1-y)^2 y+z+y (-7+6y) z+(-1+6 y) z^2\right)\right)\right)   \nonumber \\
& \qquad \qquad \qquad \qquad  \qquad \qquad \qquad \qquad\qquad\qquad \times\left( \log(1-y) \log(y)+\text{Li}_{2}(y)\right)  \nonumber \\ 
& +\frac{1}{2}  \left(\frac{2 x^2}{y^2}-\frac{2 x (1-y)}{y^2}-\frac{2(1-y) y}{x^2}+\frac{-1+2 y}{x}+\frac{-1+2 y}{y}\right)  \nonumber \\
& \qquad \qquad \qquad \qquad  \qquad \qquad\qquad \qquad\qquad \times\left( \log(1-z) \log(z)+\text{Li}_{2}(z)\right) \,,
\end{align}

\begin{align}
 C_{\beta_1}(x,y,z) = &\frac{1}{3}  \left(1-\frac{1}{y}\right) \left( 3 i \pi -\log(x)-\log(y)-\log(z)\right)  \,,\\
 C_{\beta_2}(x,y,z) = & \frac{1}{3}  \left(1-\frac{1}{z} \right) \left( 3 i \pi -\log(x)-\log(y)-\log(z)\right) \,, \\
 C_{\beta_3}(x,y,z) = & \frac{2}{3} \left( 3 i \pi -\log(x)-\log(y)-\log(z)\right) \,.
\end{align}

\subsection{$V \to gg\gamma $ at two loops}
The coefficients for the $(g_1^+,g_2^+,\gamma_3^+)$ helicity configuration are as follows:
\begin{align}
A_{\theta_1}(x,y,z)= & -\frac{2}{81} \left(81-\frac{81}{1-x}+\frac{8 (1-y)}{z^2} 
\left(-x-\frac{z^3}{x^2}+\frac{z^3}{y^2}-\frac{(1-z) z^3}{y^3}\right)\right) \nonumber \\ 
&-\frac{22}{3} \left(1-\frac{1}{y}\right)  i \pi +\frac{1}{6}  \left(-\frac{2 (1-y)}{x y}-\frac{2 (1-y)}{y z}+\frac{(-9+5 y) z}{y^3}+\frac{9 z^2}{y^3}\right) \pi ^2 \nonumber \\ 
&+\frac{1}{3}  \left(22+\frac{6}{(1-x)^2}-\frac{6}{1-x}-\frac{27 x+22 y}{y^2}\right)  \log(x) \nonumber \\ 
&-9  \frac{1}{y}  \log(y) -9  \frac{z}{y^2}   \log(z)-2  \left(\frac{z}{x y}\right)  \log(y) \log(z) \nonumber \\ 
&-2  \left(\frac{2}{y}+\frac{x}{y z}\right)  \log(x) \log(y) - \left(\frac{9x z}{y^3}+\frac{4 z}{y^2}\right)  \log(x) \log(z) \nonumber \\ 
&- \left(\frac{2}{z}+\frac{5z}{y^2}-\frac{9 (1-z) z}{y^3}-\frac{2 (1+z)}{y z}\right) \left( \log(1-x) \log(x)+\text{Li}_{2}(x)\right)  \nonumber \\ 
&+2  \left(\frac{(1-y)^2}{x y z}\right)\left( \log(1-y) \log(y)+\text{Li}_{2}(y)\right)  \nonumber \\ 
&- \left(-\frac{9 x z}{y^3}-\frac{4 z}{y^2}-\frac{2 z}{x y}\right) \left( \log(1-z) \log(z)+\text{Li}_{2}(z)\right)  \,, 
\end{align}

\begin{align}
A_{\theta_2}(x,y,z)= &-\frac{2}{81}  \left(-\frac{8 y}{x}-\frac{8 y^2}{x^2}+x^2 
\left(-\frac{8}{y^2}-\frac{8 y}{z^3}\right)+\frac{8 x \left(-1-\frac{y^3}{z^3}\right)}{y}-\frac{81y}{(1-x) z}\right)  \nonumber \\ 
&-\frac{1}{6}  \left(4+\frac{9 x^2}{y^2}+\frac{12 x}{y}+\frac{8y}{x}+\frac{8 y^2}{x^2}-\frac{-1+(8-9 x) x}{z^2}+\frac{2 (-1+6 x)}{z}\right)  \pi ^2 \nonumber \\ 
&-\frac{1}{3}  \left(\frac{46 x}{z}+\frac{27x^2}{y z}+\frac{2 (20+23 (-2+x) x) y}{(1-x)^2 z}\right)  \log(x)  -\frac{22}{3} \left(1-\frac{1}{z}\right)  i \pi \nonumber \\ 
& + \left(-\frac{9 x}{z}-\frac{8 y}{z}-\frac{8 y^2}{x z}\right)  \log(y)- \left(8+\frac{9 x}{y}+\frac{8 y}{x}\right)  \log(z) \nonumber \\ 
&+ \left(\frac{2(-2+x) x}{z^2}+\frac{10 x y}{z^2}-\frac{y^2}{z^2}\right)  \log(x) \log(y) \nonumber \\ 
&- \left(1-\frac{3x (-4+x+4 z)}{y^2}\right)  \log(x) \log(z) \nonumber \\ 
&-2  \left(\frac{(x+2 y)^2}{x^2}\right)  \log(y) \log(z) \nonumber \\ 
&+ \left(2-\frac{-1+(8-9x) x}{z^2}-\frac{2-12 x}{z}-\frac{3 x (-4+x+4 z)}{y^2}\right)   \nonumber \\
& \qquad \qquad \qquad \qquad  \qquad \qquad\qquad \qquad  \qquad \qquad \times \left( \log(1-x) \log(x)+\text{Li}_{2}(x)\right)  \nonumber \\ 
&+ \left(\frac{8 y (1-z)}{x^2}+\frac{2+y(-10+9 y)}{z^2}+\frac{6 y}{z}\right) \left( \log(1-y) \log(y)+\text{Li}_{2}(y)\right)  \nonumber \\ 
&+ \left(\frac{9 x}{y^2}+\frac{3}{y}+\frac{8 y}{x^2}-\frac{9x z}{y^2}
-\frac{3 z}{y}-\frac{8 y z}{x^2}\right) \left( \log(1-z) \log(z)+\text{Li}_{2}(z)\right) \,, 
\end{align}

\begin{align}
A_{\theta_3}(x,y,z)= & -\frac{2}{81}  \left(81+\frac{16 
\left(-y^4-y z^3+(1-z) z^3+y^3 \left(1-z+\frac{z^3}{x^2}\right)\right)}{y^2 z^2}\right) -\frac{44}{3} i \pi \nonumber  \\ 
&+\frac{1}{6}  \left(2+\frac{2}{y}+\frac{8 (1-y) y}{x^2}-\frac{-2+8 y}{x}+\frac{9 (1-y) y}{z^2}+\frac{2-6 y}{z} \right. \nonumber \\
& \qquad \left. +\frac{(9-6 y) z}{y^2}-\frac{9z^2}{y^2}\right)  \pi ^2 +\frac{1}{3}  \left(68-\frac{6}{1-x}+\frac{27 (1-x) x}{y z}\right)  \log(x)\nonumber \\ 
& + \left(9+\frac{8y}{x}+\frac{9 y}{z}\right)  \log(y)+ \left(\frac{9 (1-x)}{y}+\frac{8 z}{x}\right)  \log(z) \nonumber \\ 
&+ \left(-\frac{9 x^2}{z^2}+\frac{x (9-12 z)}{z^2}-\frac{-5+z}{z}\right)  \log(x)\log(y) \nonumber \\ 
&+ \left(-\frac{9 x^2}{y^2}+\frac{x (9-12 y)}{y^2}-\frac{-5+y}{y}\right) \log(x) \log(z) \nonumber \\ 
&-2  \left(1-\frac{4 (1-y) y}{x^2}+\frac{-1+4 y}{x}\right)  \log(y) \log(z) \nonumber \\ 
&+ \left(2+\frac{(1-x)\left(-9 (1-x)^2 x+5 (1-x) (-1+6 x) y+5 (1-6 x) y^2\right)}{y^2 z^2}\right)  \nonumber \\
& \qquad \qquad \qquad \qquad  \qquad \qquad \qquad  \qquad \qquad\times\left( \log(1-x) \log(x)+\text{Li}_{2}(x)\right)  \nonumber \\ 
&- \left(\frac{2-3x}{x}+\frac{9 (1-x) x}{z^2}+\frac{5-12 x}{z}+\frac{8 (1-x) z}{x^2}-\frac{8 z^2}{x^2}\right)  \nonumber \\
& \qquad \qquad \qquad \qquad  \qquad \qquad \qquad  \qquad \qquad\times\left( \log(1-y) \log(y)+\text{Li}_{2}(y)\right)  \nonumber \\ 
&- \left(\frac{2-3x}{x}+\frac{9 (1-x) x}{y^2}+\frac{5-12 x}{y}+\frac{8 (1-x) y}{x^2}-\frac{8 y^2}{x^2}\right)  \nonumber \\
& \qquad \qquad \qquad \qquad  \qquad \qquad \qquad  \qquad \qquad\times\left( \log(1-z) \log(z)+\text{Li}_{2}(z)\right) \,,  
\end{align}

\begin{align}
B_{\theta_i}(x,y,z)= 2 B_{\beta_i}(x,y,z)\quad \text{for} \, i=1,2,3 \,,
\end{align}

\begin{align}
C_{\theta_1}(x,y,z)= &\frac{4}{3}  \left(1-\frac{1}{y}\right) \left( i \pi -\log(x)\right) \,,\\
C_{\theta_2}(x,y,z)= &\frac{4}{3}  \left(1-\frac{1}{z}\right) \left( i \pi -\log(x)\right)  \,,\\
C_{\theta_3}(x,y,z)= &\frac{8}{3} \left(i \pi  -\log(x)\right)\,.
\end{align}
The complete results, including the other helicity configurations, 
can again be found in the attachments of the arXiv version of this paper.

\end{document}